\documentclass[%
reprint,
superscriptaddress,
nofootinbib,
amsmath,amssymb,
aps,
prl,
]{revtex4-1}

\usepackage{graphicx} 
\usepackage{dcolumn}
\usepackage{bm}
\usepackage{braket}
\usepackage{dsfont}
\usepackage{color,colortbl}
\usepackage[table,xcdraw]{xcolor}
\usepackage[colorlinks,linkcolor=black,urlcolor=black,citecolor=black]{hyperref}
\usepackage{microtype}
\usepackage{multirow}
\usepackage{subcaption}
\usepackage{mwe}
\usepackage{booktabs}
\usepackage{caption}
\usepackage{lipsum}
\usepackage{babel,blindtext}
\usepackage{amsmath}
\usepackage[toc,page]{appendix}
\usepackage[symbol*]{footmisc}
\usepackage{float}
\begin{document}
\title{Unbiased Atomistic Predictions of Crystal Dislocation \\Dynamics using Bayesian Force Fields}

\author{Cameron J. Owen$^{*,\dagger}$}
\affiliation{Department of Chemistry and Chemical Biology, Harvard University, Cambridge, Massachusetts 02138, United States}

\author{Amirhossein D. Naghdi$^{*}$}
\affiliation{NOMATEN Centre of Excellence, National Center for Nuclear Research, 05-400 Swierk/Otwock, Poland}
\affiliation{IDEAS NCBR, ul. Chmielna 69, 00-801, Warsaw, Poland}

\author{\\Anders Johansson}
\affiliation{John A. Paulson School of Engineering and Applied Sciences, Harvard University, Cambridge, Massachusetts 02138, United States}

\author{Dario Massa}
\affiliation{NOMATEN Centre of Excellence, National Center for Nuclear Research, 05-400 Swierk/Otwock, Poland}
\affiliation{IDEAS NCBR, ul. Chmielna 69, 00-801, Warsaw, Poland}

\author{Stefanos Papanikolaou$^{\dagger}$}
\affiliation{NOMATEN Centre of Excellence, National Center for Nuclear Research, 05-400 Swierk/Otwock, Poland}

\author{Boris Kozinsky$^{\dagger}$}
\affiliation{John A. Paulson School of Engineering and Applied Sciences, Harvard University, Cambridge, Massachusetts 02138, United States}
\affiliation{Robert Bosch LLC Research and Technology Center}

\def\thefootnote{$*$}\footnotetext{These authors contributed equally.}\def\thefootnote{\arabic{footnote}}
\def\thefootnote{$\dagger$}\footnotetext{Corresponding authors\\C.J.O., E-mail: \url{cowen@g.harvard.edu}\\S.P., E-mail: \url{Stefanos.Papanikolaou@ncbj.gov.pl}\\B.K., E-mail: \url{bkoz@seas.harvard.edu}\\ }\def\thefootnote{\arabic{footnote}}

% custom commands
\newcommand\bvec{\mathbf}
\newcommand{\mathsc}[1]{{\normalfont\textsc{#1}}}
\newcommand{\AN}[1]{{\color{blue}: #1}}

\begin{abstract}
Crystal dislocation dynamics, especially at high temperatures, represents a subject where experimental phenomenological input is commonly required, and parameter-free predictions, starting from quantum methods, have been beyond reach.
This is especially true for phenomena like stacking faults and dislocation cross-slip, which are computationally intractable with methods like density functional theory, as $\sim 10^5-10^6$ atoms are required to reliably simulate such systems.
Hence, this work extends quantum-mechanical accuracy to mesoscopic molecular dynamics simulations and opens unprecedented possibilities in material design for extreme mechanical conditions with direct atomistic insight at the deformation mesoscale.
To accomplish this, we construct a Bayesian machine-learned force field (MLFF) from \textit{ab initio} quantum training data, enabling direct observations of high-temperature and high-stress dislocation dynamics in single-crystalline Cu with atomistic resolution.
In doing so, a generalizable training protocol is developed for construction of MLFFs for dislocation kinetics, with wide-ranging applicability to other single element systems and alloys.
The resulting FLARE MLFF provides excellent predictions of static bulk elastic properties, stacking fault widths and energies, dynamic evolutions and mobilities of edge and screw dislocations, as well as cross-slip energy barriers for screw dislocations, all of which are compared to available experimental measurements.
This work marks the first reliable quantitative determination of dislocation mobilities and cross-slip barriers, demonstrating a substantial advancement over previous empirical and machine learned force field models.
\end{abstract}

\maketitle

\section{Introduction}
Plastic deformation understanding in metals and alloys is required for the development of novel compounds for applications in extreme conditions, such as high temperature or/and harsh irradiation environments. 
The bottleneck has traditionally been the fact that crystal plasticity is controlled by the movement of crystal defects in ways that combine multiple length scales, starting from atomistic (angstroms), and towards the mesoscale (micrometers). 
Mobile dislocations exhibit two basic forms, edge and screw, both of which achieve a given velocity when mechanical shear is applied~\cite{Anderson2017, Chu2020, Yin2021}.
The response of dislocations to various stimuli, including temperature and stress, even when just straight and topologically trivial~\cite{papanikolaou2020lambda}, determine a whole suite of material properties, including malleability, strain hardening, and grain structure, so atomistic understanding of their dynamics is required~\cite{mura2013micromechanics}. However, in contrast to Burgers vectors that can be calculated with quantum accuracy, modeling of dislocation kinetics commonly requires the input of experimental phenomenology, thus giving rise to successful interatomic potentials for mechanical applications~\cite{eam1,eam4}. 
However, in this paper, we demonstrate in the standard case of single-crystalline Cu~\cite{read2007nanoindentation, BONNEVILLE19881989}, how to perform simulations of key dislocation-kinetics-dependent properties by using machine-learning force fields (MLFF) in a manner that is unbiased and displays near-quantum accuracy at the length-scale of millions of atoms. 

\begin{figure*}[tb]
\centering
\includegraphics[width=\textwidth]{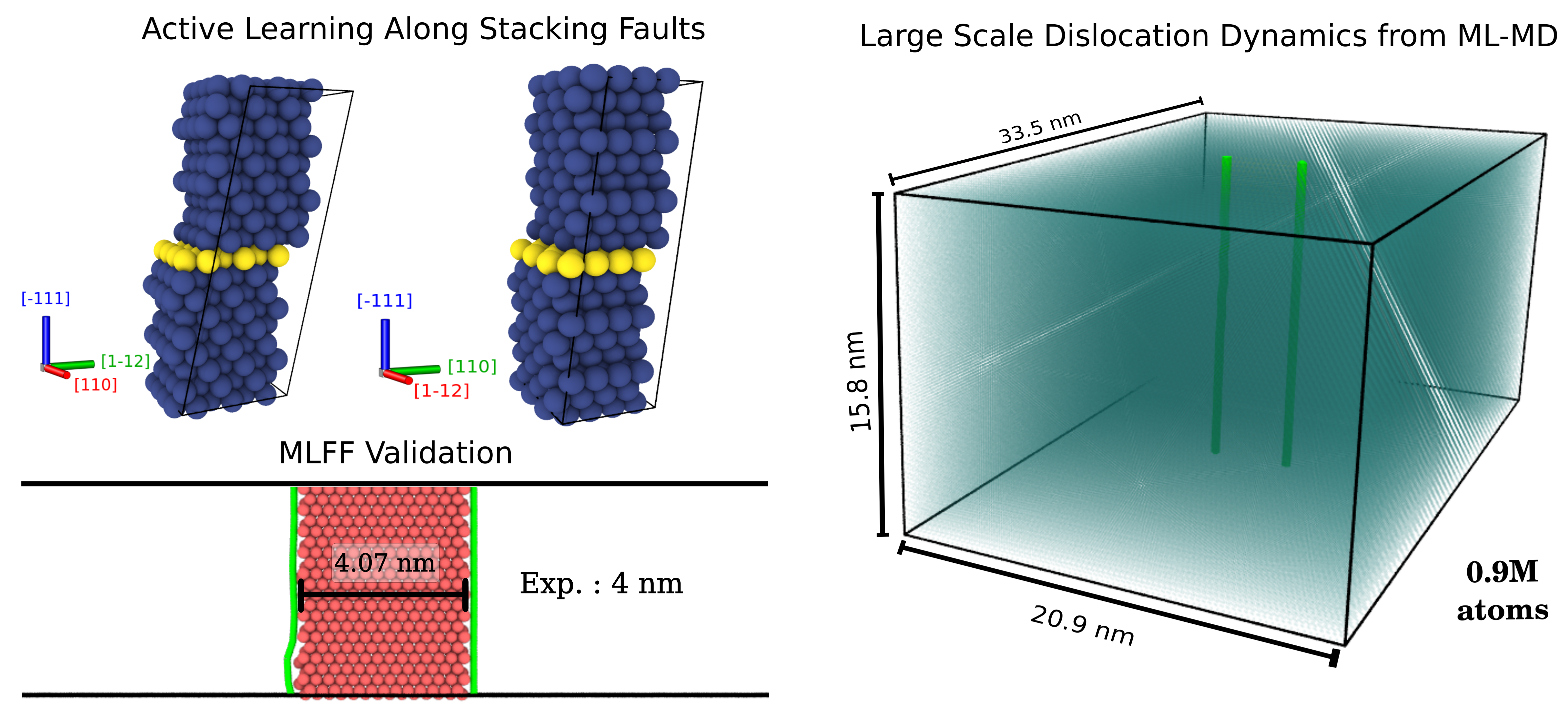}
\caption{\textbf{(Top-Left)} FLARE Active Learning is employed to construct an \textit{ab initio} training set using frames along two stacking fault directions in Cu, in addition to low- and high-temperature bulk \textit{ab initio} MD frames from the TM23 data set with vacancies \cite{owen2023complexity} 
\textbf{(Bottom-Left)} The FLARE MLFF is trained and then validated against DFT and experimental benchmarks. The MLFF is also compared to classical FFs (EAM and MEAM) for prediction of bulk properties. 
\textbf{(Right)} The MLFF is then employed in ML-MD simulations at the scale of 1-3M atoms to study edge, screw, and cross-slip dislocation dynamics as a function of temperature and mechanical stimulus. 
}
\label{fig:toc}
\end{figure*}

Dislocation dynamics may be experimentally studied with various techniques, \emph{e.g.} transmission electron microscopy (TEM) \cite{doi:10.1080/14786430600776322,ROBERTSON1999649, PhysRevLett.98.095502,BARKIA2017331,BONNEVILLE19881989}, scanning electron microscopy (SEM) \cite{STINVILLE2019152}, dark field X-ray microscopy \cite{Jakobsen:ks5610}, field ion microscopy \cite{Kim2004}, atom probes \cite{https://doi.org/10.1002/jemt.20291}, and inference microscopy (\emph{e.g.} chemical etching \cite{doi:10.1021/acs.nanolett.1c02799}). 
While these methods can provide direct structural observations of dislocations and allow for measurement of quantities like the dislocation stacking fault width (SFW) and mobility, there are inherent limitations in both length- and time-scale resolutions. 
It is these limitations that exemplify the need for atomistic insight into the dynamic evolution of dislocations under applied stimuli, naturally lending the problem to be solved using computational methods, specifically to simulate the time- and length-scales required for dislocation dynamics and cross-slip mechanisms.

Empirical force fields (FFs) are able to achieve both the time- and length-scales requirements for such simulations\cite{PhysRevB.107.094109,10.3389/fmats.2022.1046291,PhysRevResearch.4.L022043, RAO2017188, doi:10.1080/01418619908210354, doi:10.1080/01418610008212148, NOHRING201895, NOHRING2017135, OREN2017246}, but they have been shown to exhibit questionable accuracy, especially at high simulation temperatures \cite{PhysRevMaterials.7.043603} and for systems that allow for bond breaking and forming (\emph{e.g.} in catalysis \cite{Vandermause2022}).
These failure modes can be explained by the fact that empirical FFs are based on limited predefined functional forms and tuned to reproduce a small number of experimental observables (\emph{e.g.} lattice constant, melting point, densities, etc.). 
This limits their ability to extrapolate to simulation tasks in chemical and structural spaces away from the training set.
On the other hand, while \textit{ab initio} methods like density functional theory (DFT) exhibit high accuracy and transferability, they cannot capture the required length- and time-scales necessary for the study of dislocation dynamics. 
In addition, while there is a wealth of research dedicated to investigating the metallurgical aspects of crystalline materials using MLFFs, such as bulk properties, point defect formations, stacking fault energies, and more \cite{cos, curtin1, curtin2, curtin3, curtin4, curtin5}, there is a noticable scarcity of studies that specifically theoretically predict and experimentally validate dislocation mobility and cross-slip dynamics. 

In general, an empirical FF that fails to predict the GSFE landscapes would directly lead to discrepancies, such as: (1) inaccurate projection of dislocation depinning stress, which dictates the minimum applied stress required for dislocation glide, (2) artificial and incorrect predictions of the energy barriers/wells along the dislocation glide path, which can impede or elevate the dislocation velocity as well as the cross-slip energy barrier under applied stress, and (3) inaccurate dislocation interactions such as junction formations, and as a result, a less reliable description of the cross-slip mechanism(s).

Hence, to circumvent the aforementioned inaccuracies of classical FFs and limited length- and time-scales of \textit{ab initio} methods, machine-learned force fields (MLFFs) have attracted considerable attention in recent decades due to their flexible forms, ability to learn directly from high accuracy \textit{ab initio} data, and demonstrated, reliable description of mesoscopic phenomena even when limited to short-ranged descriptions of the atomic environments \cite{owen2023stability}.
Of the plethora of MLFFs that exist, the Fast Learning of Atomistic Rare Events (FLARE) code \cite{Vandermause2022,Xie2023} has yielded marked success in the description of a variety of systems from bulk materials \cite{owen2023complexity}, surfaces \cite{Lim2020EvolutionDynamics,owen2023stability}, and reactive interfaces \cite{Vandermause2022,Johansson2022Micron-scaleLearning}, to nanoparticle catalysts and their subsequent shape-change under reaction conditions \cite{owen2023unraveling}.

\begin{table*}[!htbp]
\centering
\resizebox{\textwidth}{!}{\begin{tabular}{|c|c|c|c|c|c|c|}
\multicolumn{1}{c}{\bf Ensemble}    & \multicolumn{1}{c}{\bf Direction} & \multicolumn{1}{c}{\bf Temp. (K)}   &  \multicolumn{1}{c}{\bf $\sum\tau_{\textrm{sim}}$ (ns)} & \multicolumn{1}{c}{\bf   $\sum\tau_{\textrm{wall}}$ (hr)} & \multicolumn{1}{c}{\bf $\sum N_{\textrm{DFT}}$} & \multicolumn{1}{c}{\bf $\sum N_{\textrm{runs}}$}\\
\hline
NVT      &         &            & 128.51  & 452.5  & 2162 & 13 \\ 
NPT-iso  & \{011\} & 300 - 2000 & 262.66  & 1149.5 & 5619 & 26 \\
NPT-tri. &         &            & 0.0787  & 151.7  & 851  & 24 \\
\hline
NVT      & \{1$\hat{1}$2\} & 300 - 2000 & 130.00  & 448.9  & 2080 & 13 \\ 
NPT-tri. &         &            & 0.0620  & 130.1  & 696  & 24 \\
\hline
Total & \textbf{---} & \textbf{---} & 521.3 & 2332.7 & 11408 & 100 \\
\hline
\end{tabular}}
\caption{Summary of the FLARE active learning procedure for Cu. The values provided for each system represent the sum across several independent, parallel trajectories, the total number of which is provided in the final column. Both systems employed for active learning had 24 atoms in the unit cell. These data were then combined with those for Cu from the TM23 data set in Ref. \cite{owen2023complexity} to yield the full training set, containing a total of 14,408 frames, from which the final FLARE MLFF was trained. \label{tab:active}}
\end{table*}

While other MLFFs like NequIP \cite{Batzner2021E3-EquivariantPotentials}, Allegro \cite{Musaelian2023}, MACE \cite{https://doi.org/10.48550/arxiv.2206.07697}, and MTP \cite{Shapeev2016MomentPotentials} exhibit improved accuracy relative to DFT, FLARE provides quantitative uncertainties and consequently enables an active learning workflow by which the \textit{ab initio} training set needed to yield the final MLFF can be generated with high computational efficiency, orders of magnitude faster than traditional \textit{ab initio} molecular dynamics (AIMD) sampling schemes.
The primary advantage of FLARE is its Bayesian framework based on sparse Gaussian process (SGP) kernel regression.
In practice, this means that MD during active learning is driven by the fast surrogate ML model of the potential energy, while expensive \textit{ab initio} training data is only collected during active learning when the `on-the-fly' surrogate model is uncertain of a given atomic environment. 
In contrast, traditional methods like AIMD require a quantum chemical or DFT calculation at each time step.
The set of collected training configuration snapshots (frames) can then be used to generate a MLFF for use in production MD simulations that can access both large length- and time-scales while retaining quantum mechanical accuracy.
To illustrate this ability to scale, FLARE was recently employed across 27,000 V100 GPUs on the Summit machine at Oak Ridge National Laboratory to simulate 0.5 trillion atoms for a heterogeneous catalytic reaction of H$_2$ on the Pt(111) surface \cite{Johansson2022Micron-scaleLearning}.

Specific to dislocations, MLFFs have not been used extensively for the study of their dynamics, especially at elevated temperatures and under mechanical manipulation.
However, there have been a handful of attempts using the GAP, MTP, HDNNP, and SNAP formalisms to describe crystal defects in iron \cite{zhang2023efficient,Maresca2018}, tungsten \cite{dominguezgutierrez2022atomistic}, and other transition metals and \textit{p}-block metals \cite{Freitas2022}.
However, many of these investigations, which are summarized in \cite{Freitas2022}, did not consider the dynamics of defects, including dislocations.
Hence, there is a gap in undersanding and ability to simulate dislocation dynamics at large length- and time-scales using MLFFs to provide highly accurate, direct atomistic analysis of the deformation mechanisms.

To accomplish direct observation and study of dislocation response to various stimuli using molecular dynamics, we employ active learning in FLARE to generate an \textit{ab initio} training set, here using DFT at the Perdew-Burke-Ernzerhof (PBE) approximation, from which a FLARE MLFF is constructed for the description of dislocation dynamics in Cu.
The MLFF is then validated with respect to the underlying \textit{ab initio} method and experimental quantities (\emph{e.g.} stacking fault width). 
Subsequently, we perform ML-accelerated MD (ML-MD) simulations at large time- and length scales to study explicit atomistic dislocation dynamics and cross-slip as a function of applied mechanical stress and temperature.
Ultimately, we show that the Bayesian FLARE MLFF is able to describe the generalized stacking fault energy of Cu with high accuracy relative to DFT, stacking fault width with excellent agreement to experiment, as well as edge and screw dislocation velocities and other phenomena like cross-slip mechanisms of screw dislocations with high accuracy at the scale of millions of atoms with high computational efficiency.
The complete workflow is shown schematically in Fig. \ref{fig:toc}, where all simulation targets exhibit nanosecond time-scales and mescoscopic length-scales, which makes direct comparison to experiment more realistic.
We substantiate claims of MLFF superiority over classical methods for the description of dislocation dynamics via comparison of the FLARE MLFF to existing EAM and MEAM potentials for Cu, which provide markedly worse agreement to experimental observables, in addition to providing unreliable dynamics during simulation at long time-scales.

\begin{figure*}[tb]
\centering
\includegraphics[width=\textwidth]{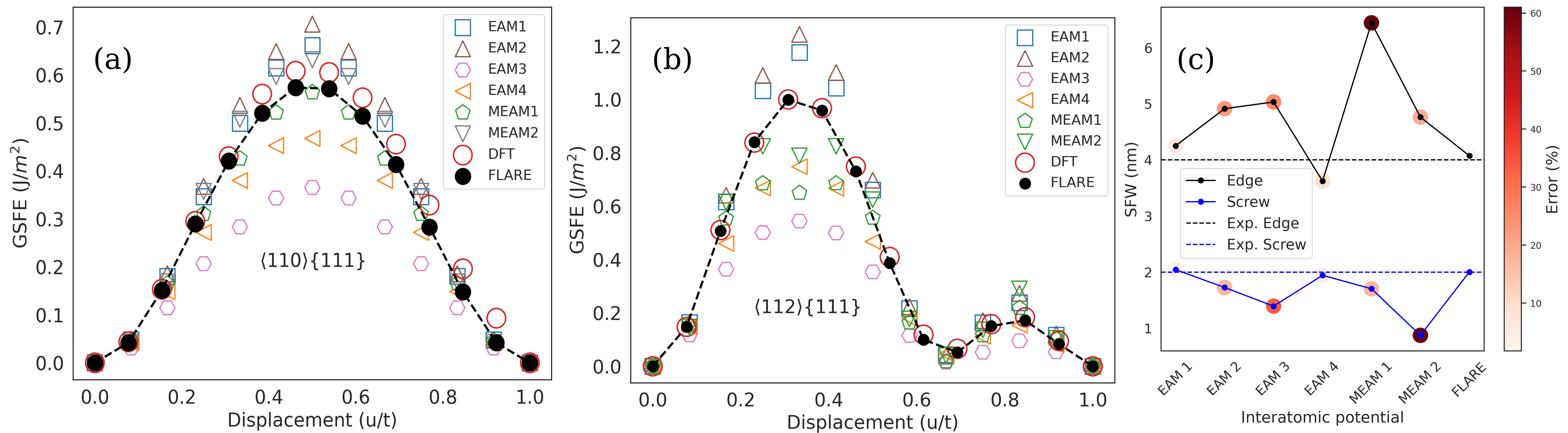}
\caption{(\textbf{a}) Generalized stacking fault energy (J/m$^2$) as a function of displacement (u/t) along the $\langle110\rangle\{111\}$ direction using classical FFs (EAM \& MEAM)~\cite{eam1,eam2,eam3,eam4,meam1,meam2}, the FLARE MLFF, and DFT. 
(\textbf{b}) Generalized stacking fault energy (J/m$^2$) as a function of displacement (u/t) along the $\langle112\rangle\{111\}$ direction using EAM, MEAM, FLARE, and DFT.
(\textbf{c}) Predictions of the stacking fault width using the same methods in panels (a) and (b), where FLARE provides the closest agreement to the experimental value of 4 nm for edge \cite{SFW1, SFW2, sfw3} and 2 nm for screw dislocations \cite{SFW1}.
}
\label{fig:gsfe}
\end{figure*}

Importantly, the FLARE ML-MD is used to determine the behavior of dislocations as a function of applied stress at high temperature, providing previously inaccessible atomistic insights into these dislocation dynamics while retaining quantum mechanical accuracy.
Specifically, we provide the most accurate predictions of stacking fault widths, dislocation mobilities under applied stress and temperature, and cross-slip rate to date, as compared to experiment, and provide atomistic insight into such phenomena, e.g. junction formation in cross-slip, and the apparent time-scales of such events. 
More fundamentally, this work provides the first instance of a semi-autonomous workflow that is able to model crystal dislocations \textit{dynamically} at appropriate length- and time-scales with \textit{ab initio} accuracy, which has not been acheived previously.

\section{Results}
\subsection{MLFF Active Learning and Validation}
\subsubsection{FLARE Active Learning is Highly Data-Efficient}
First, we provide an overview of the \textit{ab initio} reference data generation procedure, which can be applied to train MLFFs for other metals and alloys for the description of dislocation dynamics.
To start, a total of 3000 DFT frames were taken from the TM23 data set \cite{owen2023complexity}, from which an initial model for Cu was trained. 
This initial model did not provide good bulk descriptions of Cu.
This was expected, since all frames collected in the TM23 data set were from short time-scale (55 ps) NVT AIMD simulations, contained an artificially high concentration of vacancies, and only considered cells with the same number of atoms.
The TM23 data set was created explicitly for the purpose of benchmarking MLFF accuracy across transition metals rather than for reliable physical description of such systems in long time- and length-scale ML-MD.
Regardless, the TM23 data serves as a valuable basis, as it can be readily augmented with active learning using FLARE.

To do this, a set of parallel active learning trajectories were initialized using frames selected from the generalized stacking faults for Cu, specifically 13 frames along the $\langle110\rangle$ and $\langle112\rangle$ directions.
The complete set of active learning trajectories is summarized in Table \ref{tab:active}.
The FLARE code and its active learning module are described in more detail in the Methods section.
The stacking fault frames were allowed to evolve in MD simulations using a surrogate FLARE model trained `on-the-fly' in both NVT and NPT ensembles, where the latter (NPT) also included triclinic degress of freedom, where lattice vectors (x, y, and z) and angles (xy, xz, yz) are all allowed to move independently as influenced by their individual stress components, across a wide temperature (300--2000~K) and pressure ($\pm$1~GPa) ranges.
The addition of sheared cells in both the NVT, and isotropic and triclinic NPT ensembles across a broad range of temperatures and pressures allows for more diverse sampling of stress tensor components, resulting in an MLFF well suited to predict properties like bulk modulus and the elastic tensor components, as evidenced in the next section.
Ultimately, 11,408 frames were collected from the parallel active learning trajectories across the span of 76.9 hours of wall time, as all trajectories were run in parallel across 100 CPU nodes, yielding a total number of 14,408 frames when including the TM23 data.

\subsubsection{Bulk Validation}
\begin{figure}[tb]
\centering
\includegraphics[width=\columnwidth]{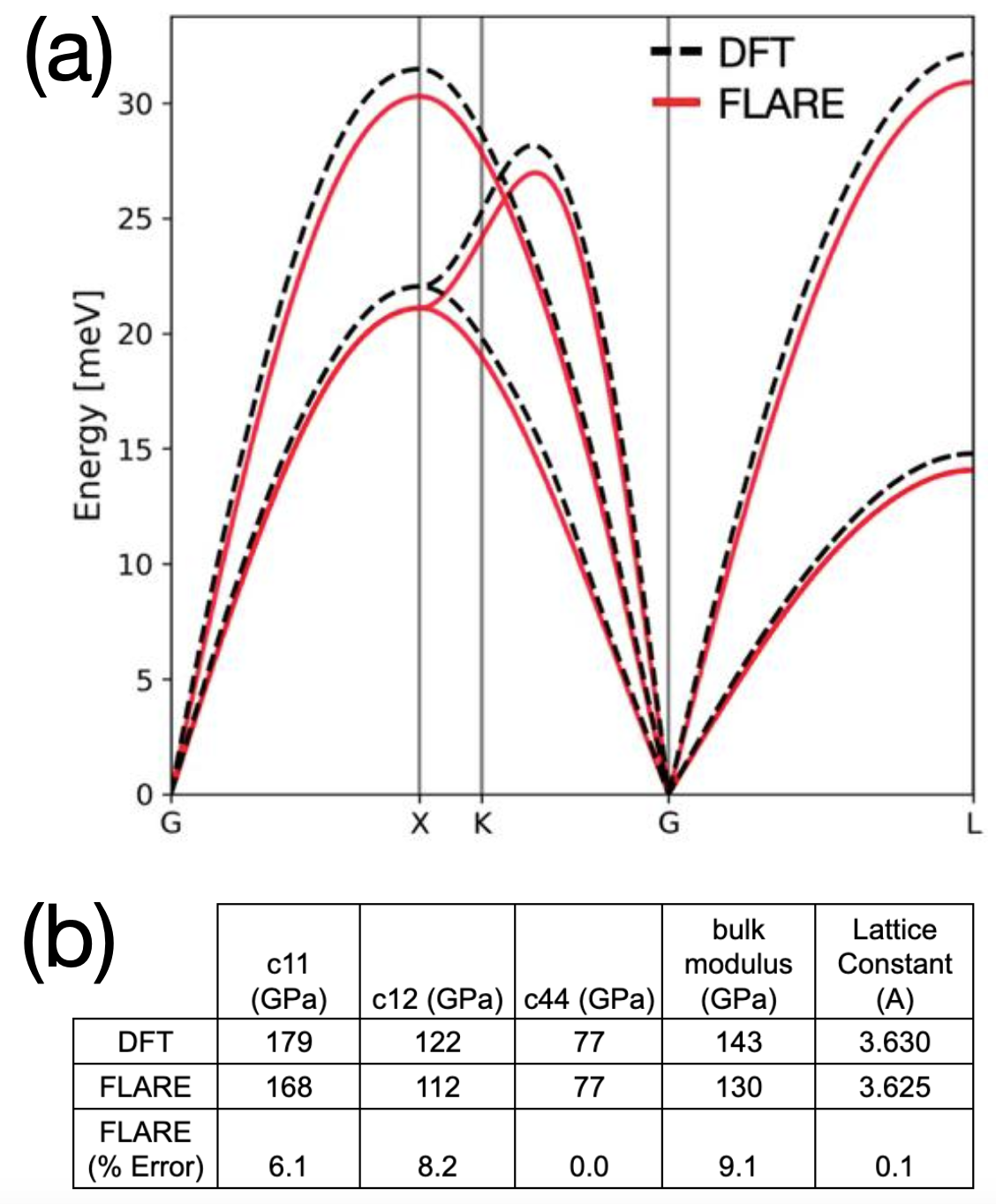}
\caption{\textbf{(a)} Phonon dispersion curves predicted by FLARE (red) and DFT (black). 
\textbf{(b)} Bulk property predictions from both DFT and FLARE model, and the absolute value of the percent error of the FLARE predictions relative to DFT.
}
\label{fig:val}
\end{figure}

Following training and energy rescaling of the FLARE MLFF, as is described in the Methods section, the model was first validated with respect to DFT for prediction of the FCC Cu lattice constant, bulk modulus, elastic tensor, and 0 K phonon dispersion. 
The lattice constant was predicted to be 3.625~\AA{}, which is in excellent agreement with the DFT predicted value of 3.63~\AA{}, and very close to the experimental value 3.6149~\AA{} \cite{Straumanis:a06566}.
The phonon dispersion results and other bulk property predictions are provided in Fig. \ref{fig:val}(a) and \ref{fig:val}(b), respectively.
The phonon dispersion curves predicted by FLARE are in excellent agreement with those predicted by DFT.
Additionally, we highlight excellent comparison of the predicted elastic tensor in panel (b), where the DFT values are computed directly here, compared to the FLARE MLFF predicted values, where most errors are less than 10 \%.

\subsection{FLARE Correctly Describes the Generalized Stacking Fault Energies and Widths}
Following validation of the FLARE MLFF against DFT for bulk property predictions, the model was then tasked with description of the stacking fault widths (SFWs) for both edge and screw dislocations and the generalized stacking fault energies (GSFEs) of Cu.
The SFW is a critical test for the MLFF, as this value effectively describes the balance between the repulsive force of two partial dislocations and the attractive force from the surface tension of the stacking fault \cite{SHANG2014168}.
Relatedly, materials with high GSFEs will deform via cross-slip or dislocation glide, whereas low GSFE materials yield larger SFWs and will not deform via cross-slip, meaning that the MLFF should be able to not only accurately describe the SFW and GSFE, but also their subsequent dynamics.
Experimental observations suggest that Cu should fall into the first category \cite{SFW1, SFW2, sfw3}, so the FLARE MLFF should yield a high GSFE, predict a small SFW in accordance with experiment, and subsequently be able to describe cross-slip as the deformation mechanism for screw dislocations if properly simulated.

The GSFE and SFW predictions across the methods are summarized in Fig.~\ref{fig:gsfe}.
In panels (a) and (b), excellent agreement is observed between the FLARE MLFF and DFT for describing the GSFE along both the $\langle110\rangle$ and $\langle112\rangle$ directions in the crystal.
As for the $\langle110\rangle\{111\}$ direction in Fig.~\ref{fig:gsfe}(a), all methods considered exhibit a maximum in the energy around 0.5 u/t, where u is the displacement and t is the lattice parameter.
In addition to the FLARE ML-MD simulations, we also considered a set of classical FFs, which yield disparate success along each GSFE curve.
Since these methods were tuned to various experimental observables, which explains their inferior performance on this explicit atomistic task.
The GSFE in Fig.~\ref{fig:gsfe}(b) along the $\langle112\rangle\{111\}$ direction, on the other hand, exhibits two maxima, the first at a displacement of 0.35 u/t and the second at 0.825 u/t.
Two peaks in the GSFE curve associated with the $\langle112\rangle\{111\}$ slip system in FCC materials result from the formation of $\frac{1}{6} \langle112\rangle$ Shockley partial dislocations along this specific crystallographic plane. This process leads to the creation of stable intrinsic stacking faults within FCC materials, which are energetically more favourable compared to a single $\frac{1}{2} \langle110\rangle$ dislocation.

Fig.~\ref{fig:gsfe}(c) summarizes the predictions of the SFWs of edge and screw dislocations in Cu, the simulation details of which are provided in the Methods section.
FLARE provides the closest agreement with respect to experiment, with the predicted SFW being 4 nm for the edge (1.75 \% error to exp. \cite{SFW1, SFW2, sfw3}) and 2 nm for the screw dislocations (in perfect agreement with exp. \cite{SFW1}). 
This observation is consistent since the GSFE plays a crucial role in determining the stacking fault width \cite{Hirth1982}. 
The SFW was not computed with the reference DFT method since the simulation cell contains 20,000 atoms and is not possible to simulate with DFT.
Of the classical FF methods tested, EAM1 and EAM4 yield the best agreement, as expected because the SFW and dislocation cores were explicitly included in their training protocols.
The empirical FFs selected in this study, specifically EAM1, EAM2, EAM3,and EAM4 are assigned to the works from Mendelev and King \cite{eam1}, Mendelev et al. \cite{eam2}, Zhou et al. \cite{eam3}, and Misin et al. \cite{eam4}, respectively, while MEAM1 and MEAM2 are assigned to the work of Etesami and Asadi \cite{meam1} and  Lee et al. \cite{meam2}, respectively.
These empirical FFs have been chosen based on their calibration to case studies closely aligned with extended defect analysis, which constitutes the primary focus of our research.

\begin{figure*}[tb]
\centering
\includegraphics[width=\textwidth]{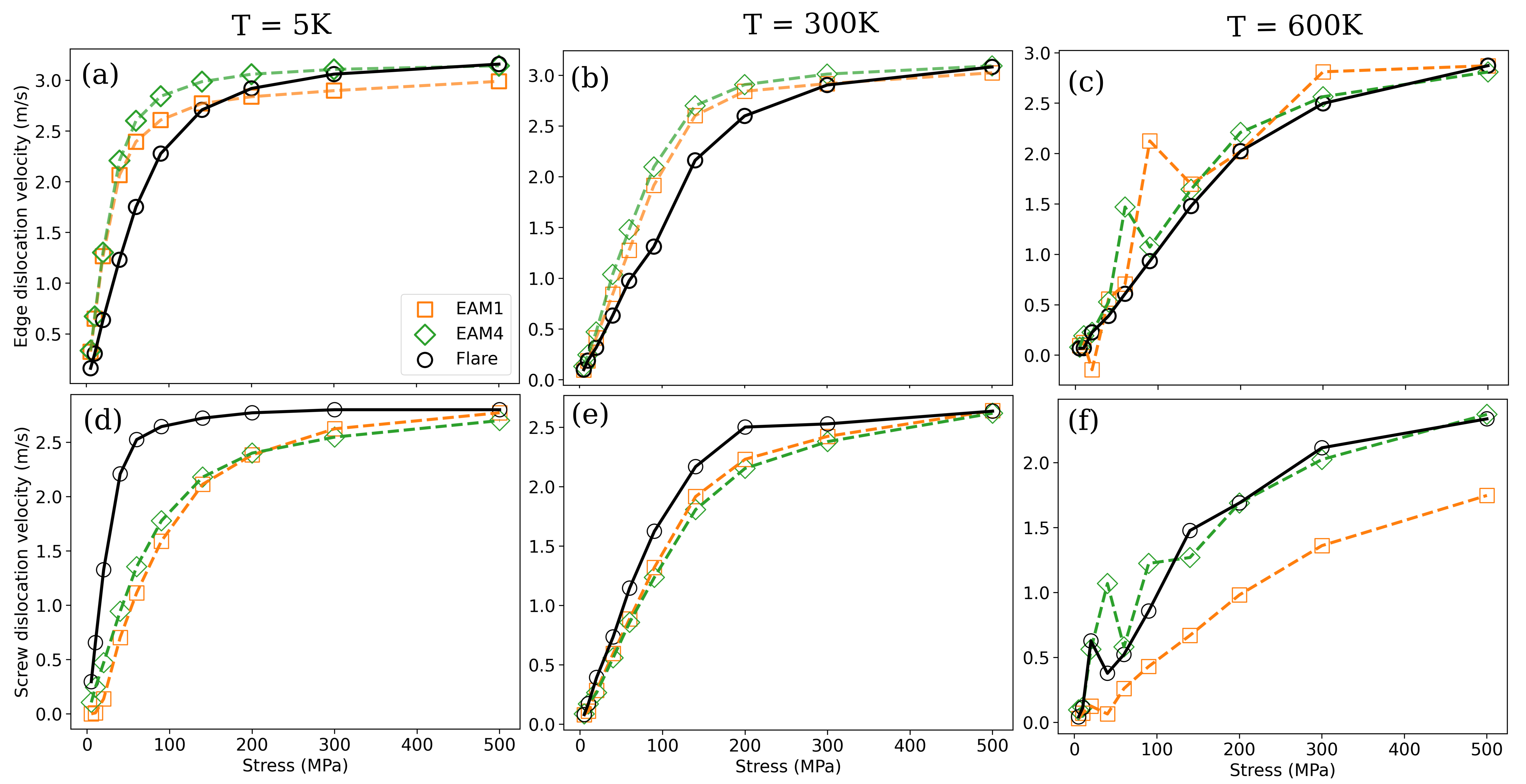}
\caption{Edge (top panel) and screw (bottom panel) dislocation velocities as a function of applied stress (MPa) and temperature (K). FLARE (black), EAM1 (orange) \cite{eam1}, and EAM4 (green) \cite{eam4} are plotted. }
\label{fig:mobility}
\end{figure*}

Crucially, we find that the FLARE MLFF predicts both GSFE curves with high accuracy relative to the underlying DFT method, which is then corroborated by accurate description of the SFWs for both dislocations of interest.

\subsection{Large-scale ML-MD Reliably Captures Dislocation Dynamics}
\subsubsection{FLARE Achieves Accurate Estimates of Mobility Coefficients for Dislocations}
Given excellent agreement between the FLARE MLFF and the reference DFT method across phonon dispersion, bulk property prediction, and GSFEs, we then deployed the model to study dislocation dynamics at various temperatures and applied mechanical stresses.
In addition to the FLARE ML-MD simulations, we also considered a set of classical FFs, in order to understand the qualitative differences between our MLFF and lower body-order empirical FFs across these stimuli, which yielded varied success in the previous task of predicting the GSFE curves and SFWs.
We selected two EAM potentials \cite{eam1, eam4} which had a lower error in prediction of SFWs (Fig.~\ref{fig:gsfe}(c)).

\begin{figure*}[tb]
\centering
\includegraphics[width=\textwidth]{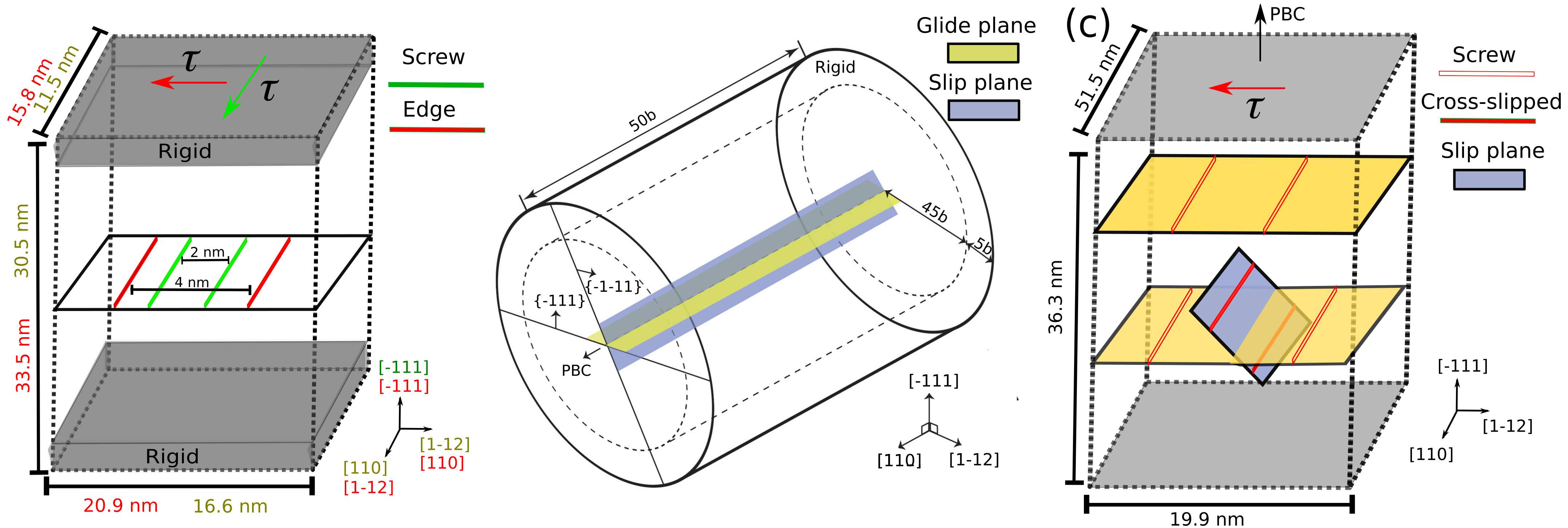}
\caption{(a) Simulation cell employed for prediction of edge and screw dislocation SFWs and dislocation mobilities.
(b) Cylindrical simulation cell employed for prediction of cross-slip NEB barrier heights.
(c) Simulation cell employed for prediction of cross-slip dynamics at 300~K and 900~MPa.
}
\label{fig:methods}
\end{figure*}

First, we considered edge and screw dislocations in Cu across a broad range of temperatures and stresses, the simulation cells of which are provided in Fig. \ref{fig:methods}. 
The simulation details are provided in the Methods section, but briefly, the procedure begins by constructing an edge or screw dislocation using Atomsk \cite{Atomsk} in an FCC Cu bulk supercell containing 0.9M atoms, which is then equilibrated for 1 ns at the desired temperature (5, 300, or 600 K) and finally simulated with various values of mechanical stress (0--500 MPa).
These results are provided in Fig.~\ref{fig:mobility}, where edge (top panel) and screw (bottom panel) dislocation velocities are observed along the axes of both temperature and applied stress.

Generally, all the FFs and FLARE considered predict a decrease in the dislocation velocity as the temperature is increased across values of applied stress.
This observation is consistent with the understanding that phonon-phonon scattering increases at higher temperatures, which can act as drag to impede dislocation movement~\cite{ma12060948}. 
This is an appropriate conclusion regarding our FLARE model given the accurate prediction of phonon dispersion by the MLFF with respect to DFT.
All FFs predict the dislocation saturation velocities at high stress (500 MPa) to essentially be equivalent (3.0 m/s for edge or 2.7 m/s for screw), except for EAM1 \cite{eam1} which predicts a screw dislocation velocity value of 1.7 m/s at T = 600K.

Focusing in more detail on the edge dislocation velocities in Fig.~\ref{fig:mobility}(a-c), we can observe that FLARE predicts smaller dislocation velocities at lower values of the applied stress, as observed by the difference in slope from 5-100 MPa.
We connect this via the experimental mobility coefficient values presented in \cite{exp_mob}, which is defined as
\begin{equation}\label{eqn:mob}
v = v_0 \left(\frac{\tau}{\tau_{0}}\right)^n\,,
\end{equation}
where $v$ and $v_0$ are the dislocation velocities at corresponding applied shear stress of $\tau$ and $\tau_{0}$, respectively. 

\begin{figure*}[tb]
\centering
\includegraphics[width=0.9\textwidth]{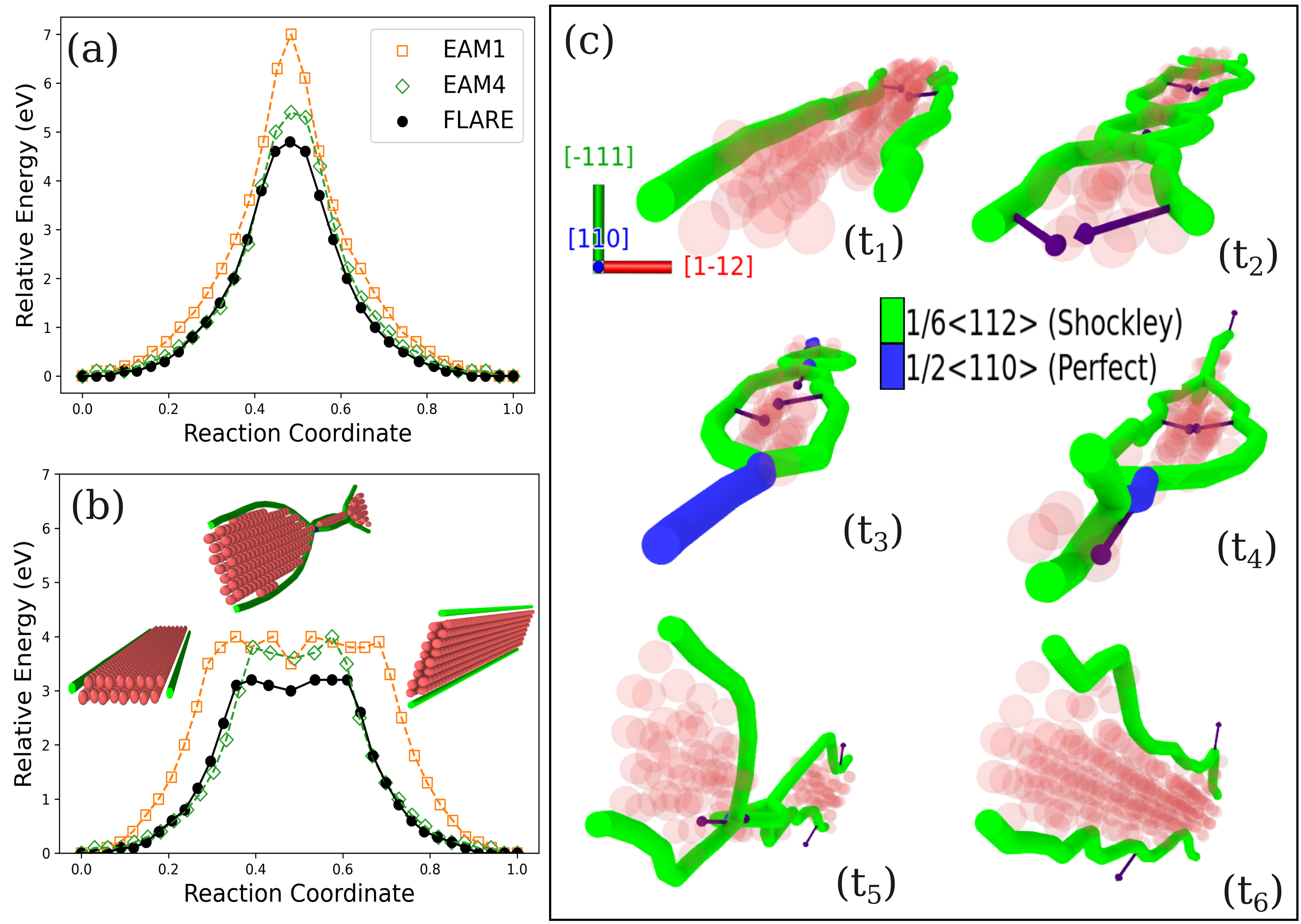}
\caption{NEB calculations of a screw dislocation cross-slip for (\textbf{a}) a short relaxation time and (\textbf{b}) a longer relaxation time. (\textbf{c}) MD simulation of screw dislocation cross-slip at T = $600$ K under 900 (MPa) applied shear stress. The partial $\frac{1}{6}\langle112\rangle$ dislocations initially on the (111) plane (\textbf{$t_{1}$}) start to form junctions (\textbf{$t_{2}$}) and perfect $\frac{1}{2}\langle110\rangle$ screw dislocation (\textbf{$t_{3}$}) before crossing to another (111) plane (slip-plane) along each segment of the stacking fault (\textbf{$t_{4}$}-\textbf{$t_{5}$}) and finally reaching a fully crossed stacking fault region on a new slip-plane (\textbf{$t_{6}$}).}  
\label{fig:cross-slip}
\end{figure*}

The mobility coefficient, $n$, is then defined by setting $\tau$ = 90 MPa and  $\tau_{0}$ = 5 MPa in Eqn. \ref{eqn:mob}, both of which fall within the linear regime of the dislocation velocity plots for 300 K. 

\begin{table}
\centering
\begin{tabular}{|c|c|c|}
\hline
\textbf{Method} & \textbf{n} &
\textbf{\% Error}\\
\hline
\textbf{EAM 1} & 1.05 & 21.6 \\
\textbf{EAM 4} & 0.96 & 9.1\\
\textbf{FLARE} & 0.89 & \textbf{1.1} \\
\textbf{Exp. \cite{exp_mob}} & 0.88 & - \\
\hline
\end{tabular}
\caption{Mobility coefficient values (n) from empirical \cite{eam1, eam4}, FLARE and experiments \cite{exp_mob} at room temperature, with FLARE having the lowest error of 1.1 \%.}
\label{tab:vel}
\end{table} 

Table \ref{tab:vel} compares the mobility coefficients obtained with FLARE, classical EAM FFs \cite{eam1, eam4} and experimental results at the same temperature (300 K). 
FLARE outperforms both classical FFs, yielding a small error of 1.1 \%, while EAM1 \cite{eam1} and EAM4 \cite{eam4} provide drastically larger errors of 9.1 and 21.6 \%, respectively. 
Additionally, FLARE presents a smooth mobility curve at high temperatures for both edge and screw dislocations across the range of stimuli considered, which is in contrast with both classical FFs.

\subsubsection{Cross-Slip Dynamics are Captured Using FLARE}
In addition to the predictive ability of FLARE with respect to edge and screw dislocations using ML-MD under a variety of conditions, we also considered the cross-slip mechanism for screw dislocations in Cu.
The cross-slip mechanism is relevant for high GSFE materials, since the Burgers vector ($\vec{b}$) points parallel to the dislocation line. 
The screw dislocation can, therefore, slip into any available glide plane, resulting in non-planar movement of the defect.
This is a difficult simulation task, given the intricate energy landscape associated with the mechanism.
Moreover, such cross-slip mechanisms have proven notoriously difficult to capture using empirical methods (\emph{e.g.} EAM and MEAM \cite{eam3, meam1}).

Hence, we targeted the cross-slip mechanism for screw dislocations in Cu as a potential advantage of the FLARE MLFF, which can be directly compared to the classical methods \cite{eam1,eam4}.
These results are provided in Fig.~\ref{fig:cross-slip}, where the cross-slip energy barrier is calculated using both Nudge Elastic Band (NEB) (panels (a) \& (b)) and during high temperature dynamical simulations via an Arrhenius relationship between the rate of cross-slip formation and temperature, as is provided in more details in the Methods section. 
The NEB calculations were performed according to the procedure outlined in the Methods.
Hence, for a small relaxation time, the results are shown in Fig.~\ref{fig:cross-slip}(a), and for longer relaxation time in Fig.~\ref{fig:cross-slip}(b), with energy barriers of $4.6$ and $3$ eV for Flare, which is lower compared to EAM1 \cite{eam1} and EAM4 \cite{eam4} by $2.4$ and $0.5$ eV in the first case, and lower to both by $0.5$ eV in the second case, respectively. 
The barrier values obtained with FLARE are closer to experimental observations reported in \cite{BONNEVILLE19881989} than EAM potentials but are still overestimated by a factor of 2, which is likely due to the NEB calculations being computed at 0 K. 

\begin{figure}[tb]
\centering
\includegraphics[width=\columnwidth]{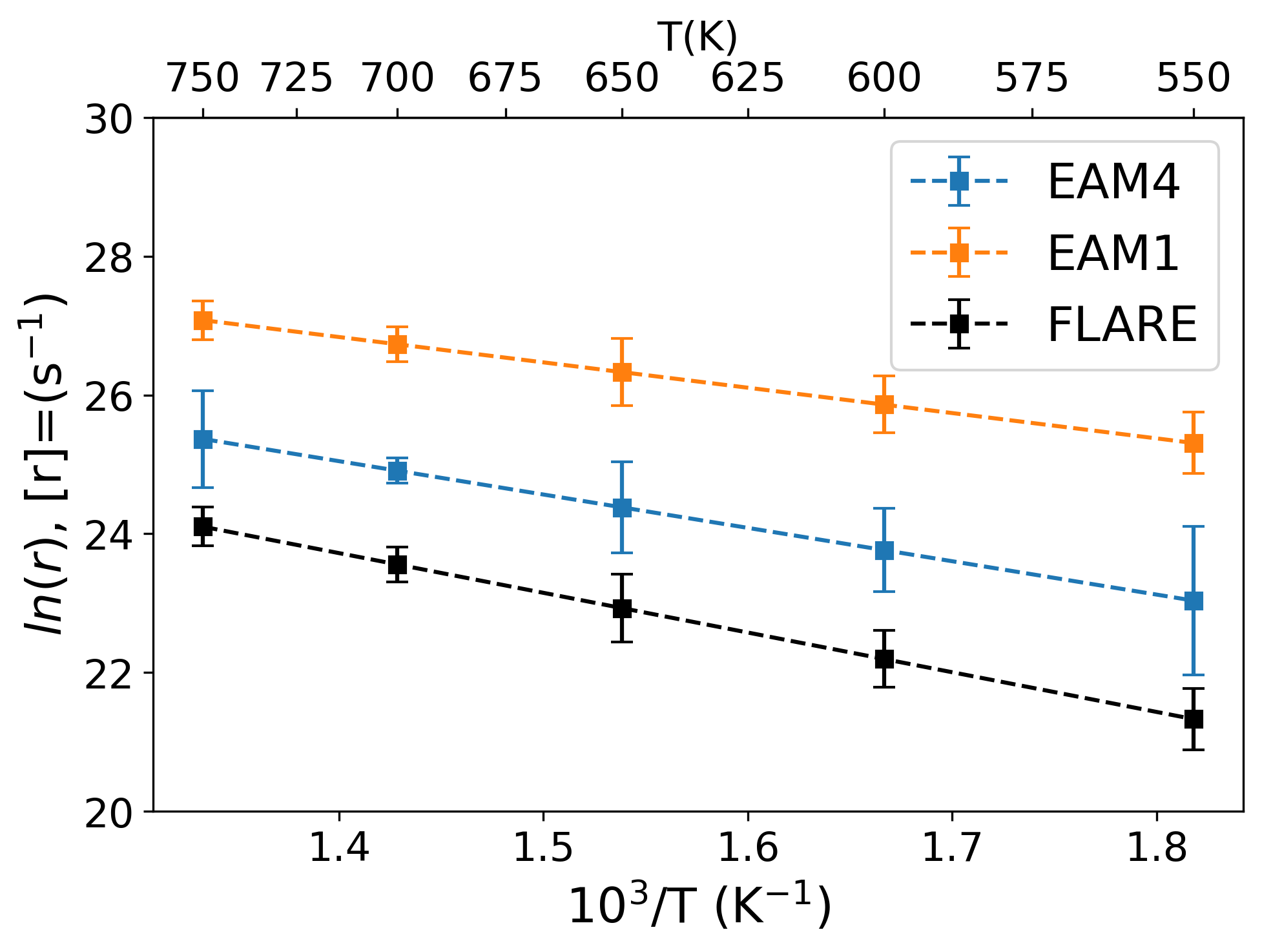}
\caption{Rate of cross-slip (in $\frac{1}{s}$) as a function of inverse temperature (in $\frac{1}{K}$) for the three methods considered, EAM1 \cite{eam1}, EAM4 \cite{eam4}, and FLARE.
}
\label{fig:arreh}
\end{figure}

High temperature (550--750 K) cross-slip MD simulations were completed by applying shear stress to the simulation box and incorporating two screw dislocations (see Methods for more details).
The cross-slip mechanism, which is well explained in \cite{OREN2017246}, is captured using FLARE. 
Fig.\ref{fig:cross-slip}(c) summarizes the cross-slip process and is demonstrated for a short screw dislocation for simplicity.
At time $t_{1}$, the dislocation is already dissociated into two $\frac{1}{6}\langle112\rangle$ partials that are first equilibrated at the target temperature. 
After the shear stress has reached the target value of $700$ MPa in 3 picoseconds (such a short simulation time is selected to prevent the cross-slip happening at the begining of the calculation), dislocation junctions are created (shown in $t_{2}$) which then are unified into perfect $\frac{1}{2}\langle110\rangle$ dislocations in $t_{3}$. 
The moment in which the first portion of dislocations are jogged is when the cross-slip mechanism is activated ($t_{4}$). 
Here, the partials disconnect, and orient themselves on a new glide plane relative to the initial plane from $t_{1}$. 
The volume of the cross-sliped portion of dislocation line, observed at $t_{4}$, is known as the activation volume \cite{HULL2011205}, which is defined as $V^{*} = L \cdot d\textbf{b}$. 
Here, $L$ is the length of cross-sliped dislocation, $d$ the average width of the newly dissociated dislocation and $\textbf{b}$ the Burgers vector value.
After both dislocation segments are moved into a second family of $\{111\}$ planes (the new slip-plane) between $t_{4}$ and $t_{6}$, the partials create a uniform stacking area along the dislocation lines. 
The time for the initiation of cross slip ($t_{4}$) is subsequently employed to determine the cross slip rate.

Cross-slip is a thermally assisted process. 
To calculate the thermal part of the activation energy, $6$ simulations at each temperature, between $550$ and $750$ K with a temperature interval of $50$ K, were performed for each potential. 
The obtained data were then fit to an Arrehnius relation of form Eqn.\ref{eq:arr}, the results of which are shown in Fig.\ref{fig:arreh}.
The cross-slip process captured by the selected EAM potentials (EAM4 and EAM1) and FLARE have the activation energies of 
$0.41$, $0.31$ and $0.49$ eV, respectively. 
The mechanical part of the activation energy barrier could be calculated as $\tau V^{*}$, where $\tau$ is the applied stress. 
The total activation energy barrier is then computed as\cite{HULL2011205}:

\begin{equation}
    \Delta E = \Delta E^{thermal} - \tau V^{*}
\label{eq:barrier}
\end{equation}

The activation volume for EAM1, EAM4 and FLARE is obtained to be $4 - 11.7$ $b^{3}$, $4.7 - 10.3$ $b^{3}$ and $5.4 - 12.3$ $b^{3}$, respectively.  
By insertion of these values to Eqn.\ref{eq:barrier}, the activation energy barriers ($\Delta E$) for EAM1, EAM4 and FLARE become $0.87 \pm 0.10$, $1.03 \pm 0.08$, and $1.13 \pm 0.03$ eV, respectively. 
Ultimately, FLARE is able to observe the cross-slip mechanism for non-planar movement of screw dislocations, yielding precise predictions for the cross-slip energy barrier compared to the experimental value of $1.15 \pm 0.37$ reported in Ref.~\cite{BONNEVILLE19881989} and the theoretical calculations done by Rao et. al~\cite{doi:10.1080/01418619908210354}.
These results are summarized in Table \ref{tab:act}.

\begin{table}[!htbp]
\centering
\scalebox{0.775}{
\begin{tabular}{|c|c|c|c|}
\hline
 &  &  \textbf{NEB$_{short}$} &  \textbf{NEB$_{long}$}\\
\textbf{Method} & \textbf{MD (eV)} &  \textbf{(eV)} &  \textbf{(eV)}\\
\hline
\textbf{EAM4} & $1.03 \pm 0.08$ & 5.1 & 3.5\\
\textbf{EAM1} & $0.87 \pm 0.10$ & 7 & 3.5 \\
\textbf{FLARE} & $1.13 \pm 0.03$ & 4.6 & 3 \\
\textbf{Exp. \cite{BONNEVILLE19881989}} & $1.15 \pm 0.37$ & -- & --\\
\hline
\end{tabular}
}
\caption{Cross-slip energy barriers ($\Delta E$) predicted by EAM \cite{eam1, eam4}, FLARE and experiments \cite{BONNEVILLE19881989} from both Arrhenius, and NEB calculations for the simulation methods.}
\label{tab:act}
\end{table} 

Despite close agreement between FLARE and the classical methods, the critical advantage of FLARE compared to EAM1 \cite{eam1} and EAM4 \cite{eam4} is the parameter-free nature of the predictions, as well as the combined closer prediction of the cross-slip energy barrier via NEB calculations at 0 K, among the previous advantages listed in the aforementioned sections for GSFE, SFW, and dislocation mobility predictions.

\subsection{Large Length-Scale Measurements from Small \textit{ab initio} Data}
Finally, we note the fact that these large length-scale observations with respect to dislocation dynamics are made despite the training data containing small \textit{ab initio} supercells ($\sim$ 20 - 80 atoms). 
This is in accordance with our prior work~\cite{owen2023stability}, where FLARE Gaussian process models based on inner-product kernels with ACE B2 descriptors were also able to capture mesoscopic surface reconstructions from very small DFT cells of Au bulk and low-index surfaces. 
This provides yet more evidence that the expressive power of small \textit{ab initio} frames, in terms of the numbers of atoms and spatial extent, and descriptors, with hard cutoff radii, are able to describe long length-scale dynamics of defects in crystals. 

\section{Discussion}
Material design for extreme conditions requires the ability to trust computational treatments of dislocation kinetics. 
In this work, a major step has been made in demonstrating how dislocation kinetics can be reliably simulated using our FLARE MLFF in ML-MD, with no parameter biases, using training datasets  comprised of cells with solely 24--71 atoms.
We showed that despite the MLFF being built on descriptions of atomic environments with a cutoff radius of 5 \AA{}, dislocation dynamics can be reliably predicted in ways that extend to much longer length-scales (10s of nanometers, 1--3M atoms).
This sets the stage for future investigations of these defects interacting with large length-scale structures, e.g. grain boundaries.
In this way, a variety of metallurgical applications may be treated in the future, including the generation of training sets for MLFFs tasked with descriptions of solute-defect interactions and/or grain-boundaries.

Most importantly, we have demonstrated the power of this unbiased active learning and simulation technique for the actual observation of dislocation dynamics, which has not been accomplished by other methods.
Moreover, we prove that our observations are in excellent agreement with experimental measurements.
Crucially, these results demonstrate the importance of our method in being able to simulate the Arrhenius behavior of cross-slip at large length-scales.
Since NEB barrier estimates are unreliable, the only direct way to assess these properties is through MD simulation.

We also introduced a semi-autonomous and generalizable training scheme that may be adopted for the quick construction of MLFFs for dislocation dynamics, through active learning along generalized  stacking fault configurations~\cite{Freitas2022}. 
The inclusion of active learning frames along idealized stacking faults under a variety of ensembles leads to reliable, unbiased modeling, capable of simulating dislocation kinetics.
Hence, we propose this training scheme as a more general, active learning, solution to training reliable MLFFs for description of dislocation dynamics.

In conclusion, we developed a FLARE MLFF that comprehensively predicts dislocation-related observables with better accuracy than available empirical methods that are commonly hand-tuned to reproduce phenomenologically consistent predictions of dislocation kinetics. 
This accomplishment is first made apparent in the prediction of the stacking fault width of Cu, where FLARE is in much better agreement with experiment than all other tested methods. 
As a result, FLARE provides a much better prediction of the stacking fault widths and generalized stacking fault energy curves, which manifests in a dramatically improved description of edge and screw dislocation mobilities, as well as the ability to capture cross-slip mechanisms, each of which provide excellent comparisons to experimental measurements. 
We believe that the accelerated autonomous simulation workflow presented here will allow practitioners to better understand the dislocation dynamics in metallic alloys with atomistic precision and near quantum mechanical accuracy, and ultimately allow for accelerated materials design for a variety of metallurgical applications.

\section{Methods}
\subsection{FLARE Active Learning}
The FLARE code was employed to construct the \textit{ab initio} (first principles) data set from which the final FLARE MLFF was trained.
FLARE is open-source and available at the following repository: \url{https://github.com/mir-group/flare}.
The Bayesian active learning module and sparse Gaussian process (SGP) have been described in detail elsewhere \cite{Vandermause2020On-the-flyEvents,Vandermause2022,Xie2021BayesianStanene,Xie2023}.
Briefly, a surrogate SGP model is trained on DFT data and used to accelerate the collection of highly relevant atomic environments driven by local uncertainty unquantification.
Environments are constructed using atomic cluster expansion (ACE) descriptors \cite{Drautz2019AtomicPotentials}, which can then be compared to one another using SGP kernel regression.
These comparisons allow for data to only be added to the SGP in an `on-the-fly' manner, meaning only when the local uncertainty of a given environment passes a threshold.
Reference \cite{Xie2023} contains a schematic of the FLARE active learning framework employed here, where each simulation is initialized using a given structure and computing atomic properties using the Sparse GP model.

As the SGP model is empty at the beginning of each active learning trajectory, all atomic environments yield a high value of the quantified uncertainty, meaning that DFT is called immediately.
For this initial call, we only add a single atomic environment to the SGP in order to reduce unneccesary duplicate environments in the SGP, which can lead to issues with numerical stability, especially in the energy predictions.
The threshold for calling DFT was set to \(10^{-4}\) in all cases, which represents the uncertainty of a given atom relative to the mean uncertainty of all atoms in the system.
Consequently, if uncertainties are above this threshold, a call is made to the Vienna ab initio Simulation Package (VASP, v5.4) \cite{Kresse1993AbMetals,Kresse1996EfficientSet,Kresse1996EfficiencySet,Kresse1999}, where DFT training labels were generated for the structure. 
The DFT parameters employed for each system are provided in the next section. 

A second threshold was also set for the addition of atomic environments to the SGP, which was \(2\cdot10^{-5}\).
The magnitudes of these thresholds are important hyperparameters that need to be tested when using FLARE to perform active learning.
Specifically, if the threshold is too large, DFT will not be called when it is necessary, leading to an unreliable surragate model with insufficient training data. Conversely, the threshold can be too small, reverting the user back to the inefficient sampling of \textit{ab initio} MD with DFT called unneccesarily on highly correlated atomic frames.
So, when the thresholds are chosen correctly, depending on the system (\emph{i.e.} number of chemical elements, type of elements, etc.) this method is drastically more efficient than \textit{ab initio} MD. 
The atomic positions in the initial frame of each simulation were randomly perturbed by 0.01 \AA{}, and the timestep employed was $\Delta t = 5$~fs.

After every call to DFT, the SGP is mapped onto a faster, but equivalent, formulation \cite{Xie2021BayesianStanene}, which is then used to perform ML-MD on the system.
Atomic positions are thus updated with respect to Newtonian mechanics using the predicted forces in LAMMPS, until another atomic environment is deemed as high uncertainty by the SGP.
Both NVT and NPT Nos\'e-Hoover ensembles were employed to perform ML-MD with the surrogate model, the latter allowing for isotropic and triclinic degrees of freedom for relaxation of the lattice vectors and angles.
As is also provided in Table \ref{tab:active} in the Main text, all systems were considered in ML-MD from 300~K to 2000~K, which allows for full exploration of crystalline and amorphous atomic environments under a variety of pressures ($\pm1$~GPa).

The FLARE hyperparameters (signal variance $\sigma$, energy noise $\sigma_E$, force noise $\sigma_F$, and stress noise $\sigma_S$) were optimized during each active learning training simulation from the 10$^{\text{th}}$ up until the 20$^{\text{th}}$ DFT call.
The priors assigned to each hyperparameter were set to empirical values observed previously for bulk Cu FLARE B2 models \cite{owen2023complexity}, specifically: $3.0$, $N_{\text{atoms}}$~meV, $0.05$~eV/\AA{}, and $0.001$~eV/\AA{}$^3$, respectively.
Each optimization step was allowed to run for a total of $200$ gradient descent steps, which was sufficient for the L-BFGS-B optimization method to achieve a tolerance of \(10^{-4}\) for the marginal log likelihood.
A total set of 26 atomic frames were used to initialize the complete set of active learning trajectories, as explained in the Main text as comprising the frames along the GSFE directions.

Following completion of all parallel active learning trajectories, the DFT frames were collected and concatenated into a master set with the set of 3000 frames from the TM23 data set \cite{owen2023complexity}, from which the final FLARE MLFF was trained. 
The TM23 data is explained in \cite{owen2023complexity}, but briefly, it is comprised of bulk frames of Cu each containing 71 atoms (each containing a single vacancy) which were taken from simulations at $0.25\cdot T_{\text{Exp. melt}}$, $0.75\cdot T_{\text{Exp. melt}}$, $1.25\cdot T_{\text{Exp. melt}}$, with each temperature yielding 1000 frames.
MLFF training followed the same atomic environment selection procedure as described above for the parallel active learning trajectories.
Specifically, environments were selected based on their relative uncertainties, but here without performing any MD, which is referred to as `offline learning.'
Subsequent to training this model, we also rescaled the energy noise hyperparameter to account for multiple systems of different sizes with respect to the number of atoms of each frame being included in the training set.
This is required, as the energy hyperparameter describes the noise level of the energy labels of the training dataset, which can easily get trapped in local minima during the optimization process, and can affect the model accuracy. 
Therefore, the energy noise hyperparameter was rescaled to 1 meV/atom, which did not influence the force or stress hyperparameters, as has been seen in other systems using the same procedue \cite{owen2023unraveling,owen2023stability}.

For both active and offline learning, the B2 ACE descriptor was employed, maintaining consistency in notation with Drautz \cite{Drautz2019AtomicPotentials}. 
By taking the atomic descriptor to the second power in the kernel, this yields ``effective'' 5-body interactions within the model, which is sufficiently complex for describing Cu with high-accuracy \cite{owen2023complexity} ($\sim 1$\% error in force, energy, and stress predictions).
Also maintaining consistency with our previous work \cite{owen2023complexity}, we employed a radial basis ($n_{\text{max}}$) of 7, angular basis ($l_{\text{max}}$) of 4, and a cutoff radius ($r_{\text{cut}}$) of 5.0~\AA{} were found to increase the log-marginal likelihood to a maximum value while reducing the energy, force, and stress errors to their respective minima.
Parity and the associated errors between the final MLFF and DFT on energy, force, and stress predictions are provided in the SI.

\subsection{Density Functional Theory}
The Vienna \textit{ab initio} Simulation Package (VASP, v5.4 \cite{Kresse1993AbMetals,Kresse1996EfficientSet,Kresse1996EfficiencySet,Kresse1999}) was employed for all DFT calculations performed within the FLARE active learning framework and subsequent validation steps. 
All calculations used the generalized gradient approximation (GGA) exchange-correlation functional of Perdew-Burke-Ernzerhof (PBE) \cite{PhysRevLett.77.3865} and projector-augmented wave (PAW) pseudopotentials. 
Semi-core corrections of the pseudopotential and spin-polarization were not included for Cu. 
A cutoff energy of $450$ eV was employed, with an artificial Methfessel-Paxton temperature \cite{PhysRevB.40.3616} of the electrons set at $0.2$ eV for smearing near the Fermi-energy. 
Brillouin-zone sampling was done using a \textbf{k}-point densities of $0.19$ \AA{}$^{-1}$ along all lattice directions centered about the $\Gamma$-point.

\subsection{MD and ML-MD Simulations}
All simulations using both FLARE and the classical FFs employed LAMMPS \cite{THOMPSON2022108171}, with ML-MD simulations using a custom pairstyle compiled for FLARE \cite{Johansson2022Micron-scaleLearning}.
GPU acceleration was achieved with the Kokkos performance portability library \cite{CARTEREDWARDS20143202}, the performance of which for FLARE has been detailed elsewhere \cite{Johansson2022Micron-scaleLearning}.
All simulations employed the Nos\'e-Hoover NPT ensemble with a timestep of $2$~fs, which is appropriate for the mass of Cu.
Velocities were randomly initialized for all simulations to a Gaussian distribution centered at the desired temperature (\emph{i.e.} 5, 300, or 600~K).
Each bulk supercell was built using the experimental lattice constant of Cu ($3.61$~\AA{}) in the Atomic Simulation Environment \cite{Larsen_2017}.

\subsubsection{GSFE}
GSFE curves were obtained by creating $1\times1 \times5$ supercells for both $\langle110\rangle\{111\}$ and $\langle112\rangle\{111\}$ slip systems in the Cu FCC crystal. 
Subsequently, the fault vectors (u) were incrementally adjusted to reach the Burgers vector ($|\vec{b}|$) value with a step size of $\delta$u=0.077$|\vec{b}|$, resulting in 13 points along the GSFE curve. 
The DFT energies were collected at each step by relaxation of the box only perpendicular to the slip directions, and subsequent FLARE calculations were completed for these relaxed geometries. 
This relaxation protocol was employed for each interatomic potential considered.
The GSFE at each point with $u\neq0$ is determined as the total energy difference of that point with the initial undefected configuration with $u=0$. 
To determine the box dimension along the slip plane, an energy convergence check with a 1~meV/atom criterion was performed. 

\subsubsection{Dislocation dynamics}

Dislocation dynamics calculations were performed for both screw and edge dislocations under applied stresses ranging from 5 to 500 MPa. 
For the $\frac{1}{2}[110](\Bar{1}11)$ edge dislocation simulations, bulk Cu supercells were generated along crystallographic directions X=$[110]$, Y=$[1\Bar{1}2]$, and Z=$[\Bar{1}11]$, with dimensions measuring $20.9\times 15.8 \times 33.5\ \text{nm}^3$, housing 901,530 atoms, as depicted in schematic Fig.~\ref{fig:methods}(a). 
Additionally, supercells incorporating the $\frac{1}{2}[1\Bar{1}2](\Bar{1}11)$ screw dislocations were generated for crystallographic orientations of X = $[1\Bar{1}2]$, Y = $[110]$ and Z = $[\Bar{1}11]$, resulting in a final box with dimensions of $16.6 \times 11.5 \times 30.5\ \text{nm}^3$, accommodating 895,860 atoms in the cell. 
Atomsk was utilized for the insertion of both edge and screw dislocations \cite{Atomsk}. 
In the case of the edge dislocation, the cell exhibited periodicity along both the dislocation glide direction (X-direction in the edge dislocation setup) and along the dislocation line (Y-direction in the edge dislocation setup). However, the screw dislocation lacks the latter form of periodicity. 
To restore periodicity along the glide path of the screw dislocation (X-direction in the screw dislocation configuration), a box vector aligned with the glide direction was tilted by an amount equal to half the Burgers vector value. 
The simulation cells feature fixed boundaries perpendicular to the dislocation line. 

For both types of dislocations, the simulation boxes were partitioned into three regions along this direction. 
The upper and lower regions, each comprising of 10~\% of the box length, contained immobilized atoms, while the remaining atoms were allowed to be mobile. 
Both dislocation types were dissociated after Conjugate Gradient (CG) relaxations \cite{CG} into two $\frac{1}{6}\langle112\rangle$ Shockley partials under the boundary conditions described. 
Regarding the screw dislocations, the atomic forces on the atoms were set to zero along the glide plane vector to ensure dissociation on the glide plane. 
After equilibration at the target temperature was reached, a shear stress was applied in the NPT ensemble to the upper region in the direction of the Burgers vector to obtain the mobility curves. 
The target applied stress is reached by increasing the force on the surface atoms by a rate of 0.01 MPa/ps so that no abrupt external pressure is experienced by the sample. 
Afterwards, the applied stress remained constant for 150 picoseconds to further obtain the mobility values.

\subsubsection{Cross-slip dynamics and NEB calculations}
Analysis of the cross-slip mechanism was done with both static Nudge Elastic Band (NEB) \cite{NEB} calculations and MD simulations. 
A schematic of the cylindrical simulation cell which was used for the NEB calculations is presented in Fig \ref{fig:methods}(b). 
After the cylindrical geometry with a radius of 20$|\vec{b}|$ was created, a perfect $\frac{1}{2}\langle110\rangle$ screw dislocation with a dimention of 40$|\vec{b}|$ along the dislocation line was inserted in the cell. 
The final dislocated cell incorporated periodic boundary conditions along the dislocation line, while an outer shell of atoms with a thickness of 5$|\vec{b}|$ (which is much larger than the interatomic potentials' cutoff radii) remained fixed during the calculations. 
The initial/cross-sliped configurations for the NEB calculations were obtained by turning the atomic forces perpendicular to the dislocation line off/on during the CG initial relaxation. 
While the outer shell of atoms were fixed, the energies along the NEB transition path were calculated by minimization of atomic positions using the FIRE algorithm~\cite{Fire} for each point along the path. 
The relaxations were done for a maximum step of 350 due to the fact that further relaxations result in unrealistic configurational states along the cross-slip mechanism~\cite{NOHRING2017135}. 
Further relaxation results in dislocation junctions, or generally configurations along the reaction path (other than the initial and fully cross-slipped one), which can settle into either the initial or final states or adopt states along new glide planes.
Hence, limiting the relaxation time permits only a single observation of cross-slip.

To simulate the cross-slip mechanism and calculate the energy barrier using MD, a fully periodic supercell with a dimention of $19.9\times 51.1 \times 36.3\ \text{nm}^3$, containing 3,132,000 atoms and two perfect screw dislocations with opposite $\vec{b}$ vectors was created (Fig.~\ref{fig:methods}(c)). 
After the simulation cell was equilibrated at the target temperature for 2~ns, an Escaig stress ($\tau$) was applied with a relatively high rate of 0.28 ~pa/fs so that the cross-slip did not occur due to aburpt external pressure. 
Afterwards, the applied stress remained constant until the cross-slip was observed. 
The cross-slip energy barrier was then calculated using an Arrhenius expression,

\begin{equation}
r(T) = \nu \exp \left( {-\frac{\delta E}{k_{\text{B}}T}} \right)    
\label{eq:arr}
\end{equation}

where $r(T)$ is the cross-slip rate at temperature $T$, $\nu$ is the rate prefactor, $\delta$E is the energy barrier and $k_{\text{B}}$ is the Boltzmann constant. 

\subsection{MD and ML-MD Simulation Post-Processing}
All visualization of the simulations was performed using OVITO~\cite{ovito}. 
The Python-API for OVITO was employed to perform analyses of the trajectories detailed above.

\section{Data Availability}
All \textit{ab initio} active learning training data, and the training scripts will be provided on the Materials Cloud upon publication. 
The TM23 data set and availability is described in \cite{owen2023complexity}.

\subsubsection*{Author Contributions}
C.J.O. augmented the TM23 dataset for Cu with FLARE active learning as described in the Results and Methods sections, compiled and analyzed the data, performed all MLFF training, validation, and FLARE ML-MD simulations. 
N.D.A. performed all classical FF MD simulations, developed simulation protocols for evaluation of each of the dislocations considered, and performed the cross-slip FLARE ML-MD simulations.
C.J.O. and N.D.A. jointly created all figures, completed post-processing of the data, and wrote the manuscript. 
A.J. provided useful advice concerning the FLARE MD simulations.
D.M. provided useful input concerning the EAM and MEAM simulations performed by N.D.A.
S.P. supervised all aspects of the dislocation simulations and interpretation of the results.
B.K. supervised all aspects of the work.
All authors contributed to revision of the manuscript.

\subsubsection*{Acknowledgements}
This work was supported primarily by the US Department of Energy, Office of Basic Energy Sciences Award No. DE-SC0022199 as well as by Robert Bosch LLC. 
C.J.O. was supported by the National Science Foundation Graduate Research Fellowship Program under Grant No. DGE1745303. 
N.D.A. and D.M. were supported by  European Union
Horizon 2020 Research and Innovation Programme under grant agreement no. 857470 and from the European
Regional Development Fund via the Foundation for Polish Science International Research Agenda PLUS program grant No. MAB PLUS/2018/8.
A.J.\ was supported by the NSF through the Harvard University Materials Research Science and Engineering Center Grant No. DMR-2011754. 
This research used resources of the National Energy Research Scientific Computing Center (NERSC), a DOE Office of Science User Facility supported by the Office of Science of the U.S. Department of Energy under Contract No. DE-AC02-05CH11231 using NERSC award BES-ERCAP0024206.
Computing resources were also provided by the FAS Division of Science Research Computing Group at Harvard University and the National Centre of Nuclear Research.

\subsubsection*{Competing Interests}
The authors declare no competing interests.

\section{References}
\bibliography{bib.bib}

\end{document}